\documentclass[letterpaper]{article} 
\usepackage{aaai25}  
\usepackage{times}  
\usepackage{helvet}  
\usepackage{courier}  
\usepackage[hyphens]{url}  
\usepackage{graphicx} 
\usepackage{natbib}  
\usepackage{caption} 
\usepackage{algorithm}
\usepackage{algorithmic}

\usepackage{newfloat}
\usepackage{listings}
\DeclareCaptionStyle{ruled}{labelfont=normalfont,labelsep=colon,strut=off} 
\floatstyle{ruled}
\newfloat{listing}{tb}{lst}{}
\floatname{listing}{Listing}
\usepackage{amsmath}

\DeclareMathOperator*{\argmin}{\,min}
\usepackage[algo2e,ruled,noend,linesnumbered]{algorithm2e}
\usepackage{amsthm}
\usepackage{algorithm} 
\usepackage{subcaption}

\usepackage{subfig}
\usepackage{bm}
\usepackage{multirow}

\newtheorem{example}{Example}

\usepackage{booktabs}
\newif\ifSpacePermits

\SpacePermitsfalse

\newcommand{\rt}{{:}}

\usepackage{color,soul}
\usepackage{xcolor}

\title{EHL*: Memory-Budgeted Indexing for Ultrafast Optimal Euclidean Pathfinding}
\author{
    Jinchun Du, Bojie Shen, Muhammad Aamir Cheema
}
\affiliations{
    Faculty of Information Technology, Monash University, Melbourne, Australia\\
    \{jinchun.du,
    bojie.shen,
    aamir.cheema\}@monash.edu
}

\usepackage{bibentry}

\begin{document}

\maketitle

\begin{abstract}
The Euclidean Shortest Path Problem (ESPP), which involves finding the shortest path in a Euclidean plane with polygonal obstacles, is a classic problem with numerous real-world applications. The current state-of-the-art solution, Euclidean Hub Labeling (EHL), offers ultra-fast query performance, outperforming existing techniques by 1-2 orders of magnitude in runtime efficiency. However, this performance comes at the cost of significant memory overhead, requiring up to tens of gigabytes of storage on large maps, which can limit its applicability in memory-constrained environments like mobile phones or smaller devices. Additionally, EHL's memory usage can only be determined after index construction, and while it provides a memory-runtime tradeoff, it does not fully optimize memory utilization.
In this work, we introduce an improved version of EHL, called EHL*, which overcomes these limitations. A key contribution of EHL* is its ability to create an index that adheres to a specified memory budget while optimizing query runtime performance. Moreover, EHL* can leverage pre-known query distributions—a common scenario in many real-world applications—to further enhance runtime efficiency. Our results show that EHL* can reduce memory usage by up to 10-20 times without much impact on query runtime performance compared to EHL, making it a highly effective solution for optimal pathfinding in memory-constrained environments.


\end{abstract}

%

\section{Introduction}


The Euclidean Shortest Path Problem (ESPP) finds the shortest obstacle-avoiding path between a given source and target for an agent to travel. ESPP is a well-studied problem with various real-world applications, including computer games~\cite{computer_games}, robotics~\cite{mac2016heuristic}, and indoor navigation~\cite{cheema2018indoor}. 
In many of these applications, it is crucial to compute the optimal shortest path as fast as possible, especially for large-scale deployments that require calculating tens of thousands of paths per second. 
This challenge has led to the development of numerous algorithms, such as navigation-mesh-based planners like Polyanya~\cite{polyanya}, enhanced variations of visibility graphs like hierarchical sparse visibility graphs~\cite{sparse_vis}, and oracle-based approaches like End Point Search (EPS)~\cite{EPS} and Euclidean Hub Labeling (EHL)~\cite{EHL}.


Among the existing optimal algorithms, the state-of-the-art is EHL, which exploits Hub Labeling~\cite{abraham2011hub}, the leading shortest path approach for graphs such as road networks. During the preprocessing phase, EHL constructs a visibility graph on the convex vertices of the polygonal obstacles and precomputes the hub labeling on this graph. The precomputed hub labels of each vertex are then copied to each grid cell of a uniform grid that is visible from the vertex. During the online phase, EHL considers the labels stored in the grid cells containing the start and target, respectively. These labels are joined to obtain the shortest path via a merge-join process.

EHL is 1-2 orders of magnitude faster than existing algorithms and performs best when grid cells are small (i.e., there are many grid cells). However, in this case, it requires substantial memory to store the labels, limiting its use in memory-constrained environments. Although EHL offers a memory-runtime trade-off by increasing the grid cell size to reduce memory usage at the cost of longer running times, it still faces two limitations: 1) Memory required by EHL cannot be predicted in advance and is only known after the index is constructed. This is problematic for smaller devices with fixed memory budgets that need to create an index within the budget while optimizing runtime. 2) In many real-world cases, queries often follow a specific distribution, with certain areas of maps receiving more queries than others. While works in other domains~\cite{sheng2023wisk,tzoumas2009workload} show that this information can be exploited to improve performance, EHL does not take advantage of this information, even when it is available.

\begin{figure}[t!]
    \centering
    \begin{subfigure}{0.5\linewidth} 
        \centering
        \includegraphics[width=\textwidth]{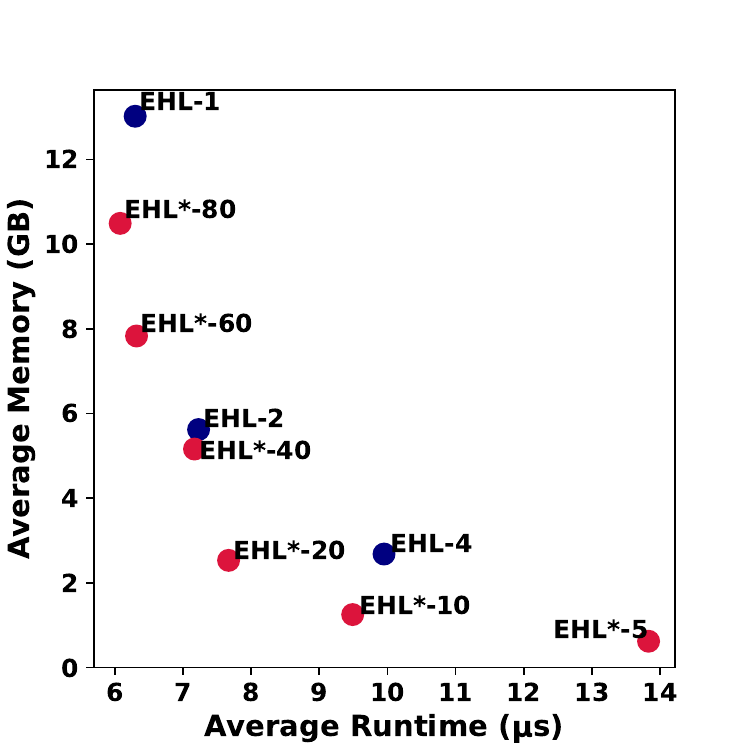}
        \caption{Unknown distribution}
        \label{fig:intro_unif}
    \end{subfigure}
    \begin{subfigure}{0.48\linewidth} 
        \centering
        \includegraphics[width=\textwidth]{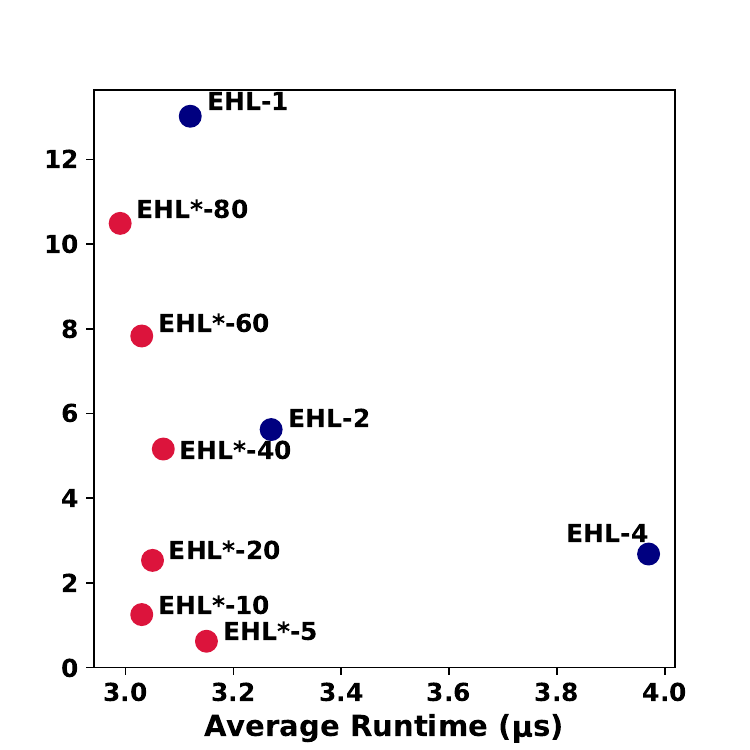}
          \caption{Known distribution}
        \label{fig:intro_clust}
    \end{subfigure}  
    \caption{Memory-runtime tradeoff provided by EHL and EHL* on the Expedition map.}
    \label{fig:intro_exp}
\end{figure}

In this paper, we address these limitations of EHL and propose EHL*, an improved version of EHL that works within a memory budget $\mathcal{B}$ to optimize the query runtime while still guaranteeing to find the optimal solution. Moreover, it is also able to exploit the expected  query distribution if it is known beforehand. 
EHL* builds on EHL by incorporating a compression phase that merges the labels of adjacent grid cells into arbitrary-shaped regions. Each merging effectively reduces the memory because the labels that were stored twice in two different cells/regions are now likely to be stored once in one larger region. The key is to carefully select which cells or regions to merge, ensuring that memory savings are achieved without significantly increasing query runtime. EHL* continues to merge chosen neighboring regions until the required memory drops below $\mathcal{B}$. 

Figure~\ref{fig:intro_exp} illustrates the memory-runtime trade-off achieved by EHL and EHL* on the Expedition map from the StarCraft benchmark~\cite{gridBenchmarks}. Following the methodology of the original work~\cite{EHL}, EHL-1 represents the algorithm with grid cells the same size as those in the benchmark. EHL-2 and EHL-4 correspond to versions where the grid cell sizes are increased by factors of 2 and 4 in each dimension, respectively. EHL*-x denotes EHL* with a memory budget $\mathcal{B}$ set to x\% of the memory used by EHL-1. 

Figure~\ref{fig:intro_unif} presents results assuming the query distribution is unknown. EHL* not only allows specifying a memory budget but also delivers a superior memory-runtime trade-off. For example, EHL*-60 reduces memory usage by 60\% compared to EHL-1 without significantly increasing runtime. Figure~\ref{fig:intro_clust} shows results when queries are assumed to be clustered in two specific areas of the map, and EHL* leverages this information to strategically merge cells/regions. The results are even more impressive, with EHL*-5 reducing memory from over 10GB to around 600MB while maintaining similar query runtime performance.

\section{Preliminaries}

An obstacle is represented by a \textbf{polygon} that consists of a closed set of edges, each associated with two points at its ends, known as vertices. A \textbf{convex vertex} (resp. \textbf{non-convex vertex}) is a vertex located at a convex (resp. concave) corner of the polygon.
Two points are said to be \textbf{visible} to each other (also known as \textbf{co-visible}) iff there exists a straight line between them that does not pass through any obstacle.
A \textbf{path} $\mathcal{P}$ between a source $s$ and target $t$ is a sequence of points $\langle p_1$,$p_2$, $\cdots$, $p_n\rangle$ where $p_1=s$,  $p_n=t$ and every successive pair of points $p_i$ and $p_{i+1}$ ($i<n$) is co-visible. 
The \textbf{length} of a path $\mathcal{P}$ is the total sum of the Euclidean distances between each consecutive pair of points along the path, denoted as $|\mathcal{P}|$, i.e., $|\mathcal{P}| = \sum _{i=1}^{n-1}Edist(p_i,p_{i+1})$ where $Edist(p_i,p_{i+1})$ is the Euclidean distance between $p_i$ and $p_{i+1}$.
The \textbf{Euclidean Shortest Path Problem (ESPP)} computes the shortest obstacle-avoiding path between a given $s$ and $t$ with the minimum length, denoted as $sp(s,t)$. The shortest distance (i.e., the length of shortest path) between $s$ and $t$, denoted as $d(s,t)$, i.e., $|sp(s,t)| = d(s,t)$.

\section{Euclidean Hub Labeling (EHL)}

This section introduces Euclidean Hub Labeling (EHL), the current state-of-the-art algorithm, which we further improve in this work. EHL consists of two phases discussed shortly: an offline preprocessing phase and an online query phase. 



\subsection{Offline Preprocessing}


\subsubsection{Building Hub Labelling on Visibility Graph}
Visibility graph is a popular concept utilized in many pathfinding algorithms for Euclidean space, e.g.,~\cite{sparse_vis,EPS}. The graph $G = (V, E)$ consists of a set of convex vertices $V$ and a set of edges $E$, where each edge $e \in E$ connects a pair of co-visible vertices in $V$. 
Given the constructed visibility graph $G = (V, E)$, Hub Labeling (HL) computes and stores, for each vertex $v_j\in V$, a set of hub labels denoted as $H(v_j)$. Each hub label in $H(v_j)$ is a tuple $(h_i,d_{ij})$ that includes: (i) a hub vertex $h_i \in V$; and (ii) the shortest distance $d_{ij}$ between the hub vertex $h_i$ and $v_j$. HL must satisfy the \emph{coverage property}~\cite{hl_np}, which means that for every pair of reachable vertices $v_j\in V$ and $v_k \in V$, $H(v_j)$ and $H(v_k)$ must contain at least one hub vertex $h_i$ on the shortest path from $v_j$ to $v_k$. Thus, the shortest distance between any $v_s\in V$ and any $v_t\in V$ can be efficiently determined as follow: 

\begin{equation}\label{eq:hl}
d(v_s,v_t) = \argmin_{h_i\in H(v_s) \cap H(v_t)} (d_{is} + d_{it})
\end{equation}
According to Eq.~\eqref{eq:hl}, the \textbf{shortest distance} between $v_s$ and $v_t$ can be calculated by performing a simple scan of the sorted label sets $H(v_s)$ and $H(v_t)$. The complexity is $O(|H(v_s)| + |H(v_t)|)$, where $|H(v_j)|$ represents the number of labels in $H(v_j)$. The shortest path can be retrieved by utilizing the successor node for each label. We refer the reader to \hbox{~\cite{shp}} for details.


\begin{example}
Figure~\ref{fig::VG} shows the visibility graph constructed for the example in Figure~\ref{fig::running_example}, containing 5 convex vertices.
Table~\ref{tab:hublabels} shows the hub labels for this visibility graph. The computation of d(E,A) involves scanning the hub labels for vertices E and A, during which two common hub nodes, B and E, are identified. Here, d(E,B) + d(B,A) = 6.1 + 5.1 = 11.2 and d(E,E) + d(E,A) = 0 + 10. Thus d(E,A) = 10.
\end{example}

\subsubsection{Computing Euclidean Hub Labelling }
With the HL constructed on the visibility graph, we can efficiently solve ESPP when both the start $s$ and target $t$ are at convex vertices. However, since $s$ and $t$ can be arbitrary locations, EHL adapts HL to handle these cases, as follows.
EHL overlays a uniform grid on the map, with grid size adjustable to manage memory usage. For each cell $c$, EHL computes:
\begin{itemize}
    \item A visibility list $L_c$: This list consists of every convex vertex $v$ such that at least some part of $c$ is visible from $v$.
    
    \item A via-labels list $VL(c)$: For each vertex $v_j$ in the visibility list $L_c$, we consider each hub label $(h_i,d_{ij})\in H(v_j)$ and insert a via-label $h_i\rt(v_j,d_{ij})$ into $VL(c)$. 
    Intuitively, this label indicates that the cell $c$ is visible from a via vertex $v_j$ and there is a potential path from $c$ to the hub node $h_i$ via $v_j$ with the distance between $h_i$ and $v_j$ being $d_{ij}$
\end{itemize}

\begin{figure}[t]
    \centering
    \begin{minipage}{0.22\textwidth}
        \centering \scriptsize
        \includegraphics[width=\textwidth]{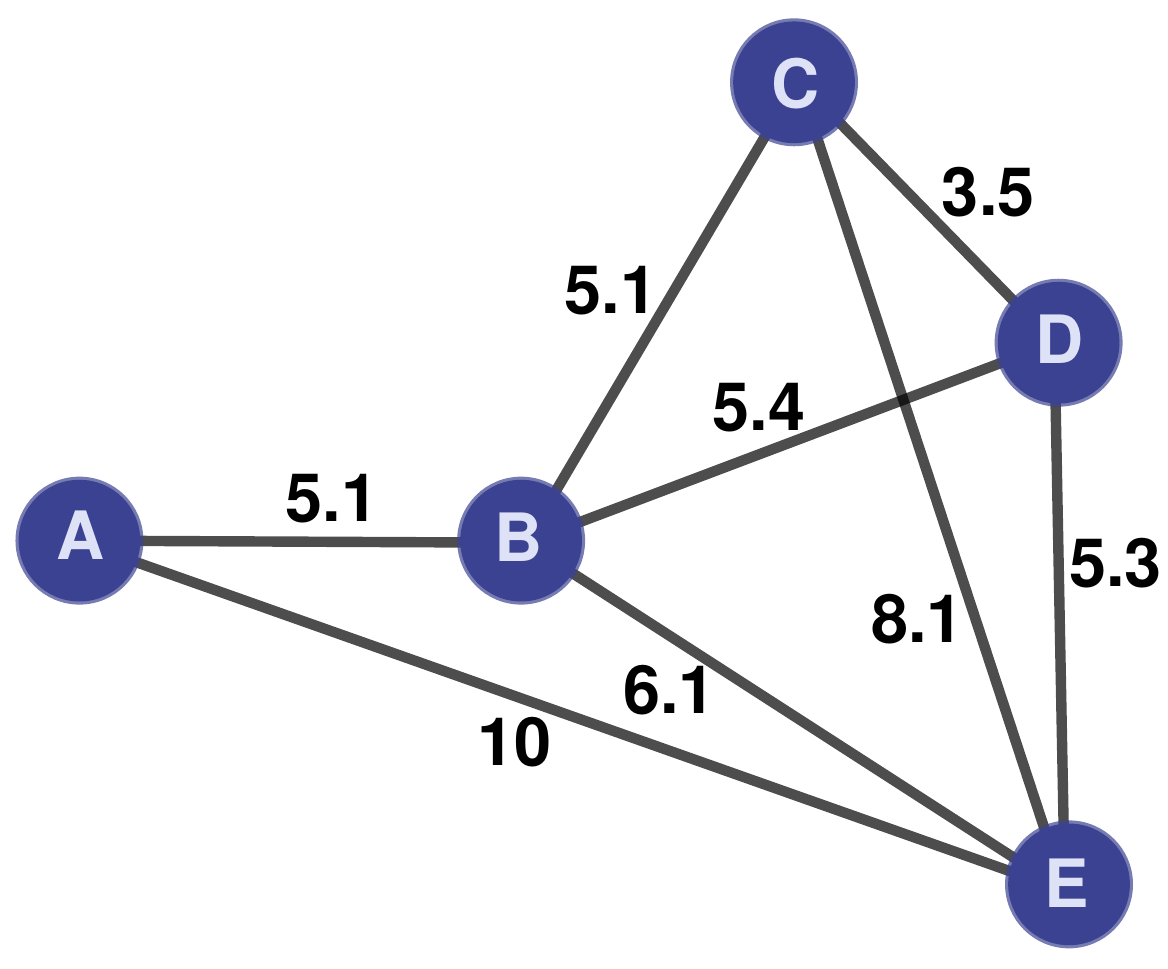} 
        \caption{Visibility graph for the example in Figure~\ref{fig::running_example}}
        \label{fig::VG}
    \end{minipage}
    \hspace{0.015\textwidth}
    \begin{minipage}{0.22\textwidth}
    \renewcommand{\arraystretch}{1.2}
        \centering \scriptsize
        \vspace{1.4\baselineskip}
        \begin{tabular}{| c | l | }
    \hline
    \textbf{Vertex}  & \textbf{Hub labels}  \\
     \hline \hline
     A & (A, 0), (B, 5.1), (E, 10)
     \\ \hline
     B & (B, 0) 
     \\ \hline
     C & (B, 5.1), (C, 0)  
     \\ \hline
      D & (B, 5.4), (D, 0) 
     \\ \hline
      E & (B, 6.1), (D, 5.3), (E, 0) 
     \\ \hline
    \end{tabular}
    \renewcommand{\arraystretch}{1.2}
        \captionof{table}{Hub labeling for the graph in Figure~\ref{fig::VG}}
        \label{tab:hublabels}
    \end{minipage}
\end{figure}

Within each cell $c$, there may be multiple via-labels containing the same hub nodes. We use $VL_{h_i}(c)$ to denote the set of all via-labels in $c$ that have the same hub node $h_i$. $H(c)$ represents the set of unique hub nodes for the labels stored in $c$. For faster query processing, $H(c)$ is sorted by hub nodes $h_i$ to efficiently join the hub labels of two different cells. 

\begin{figure}[t]
\centering
\includegraphics[width=0.8\columnwidth]{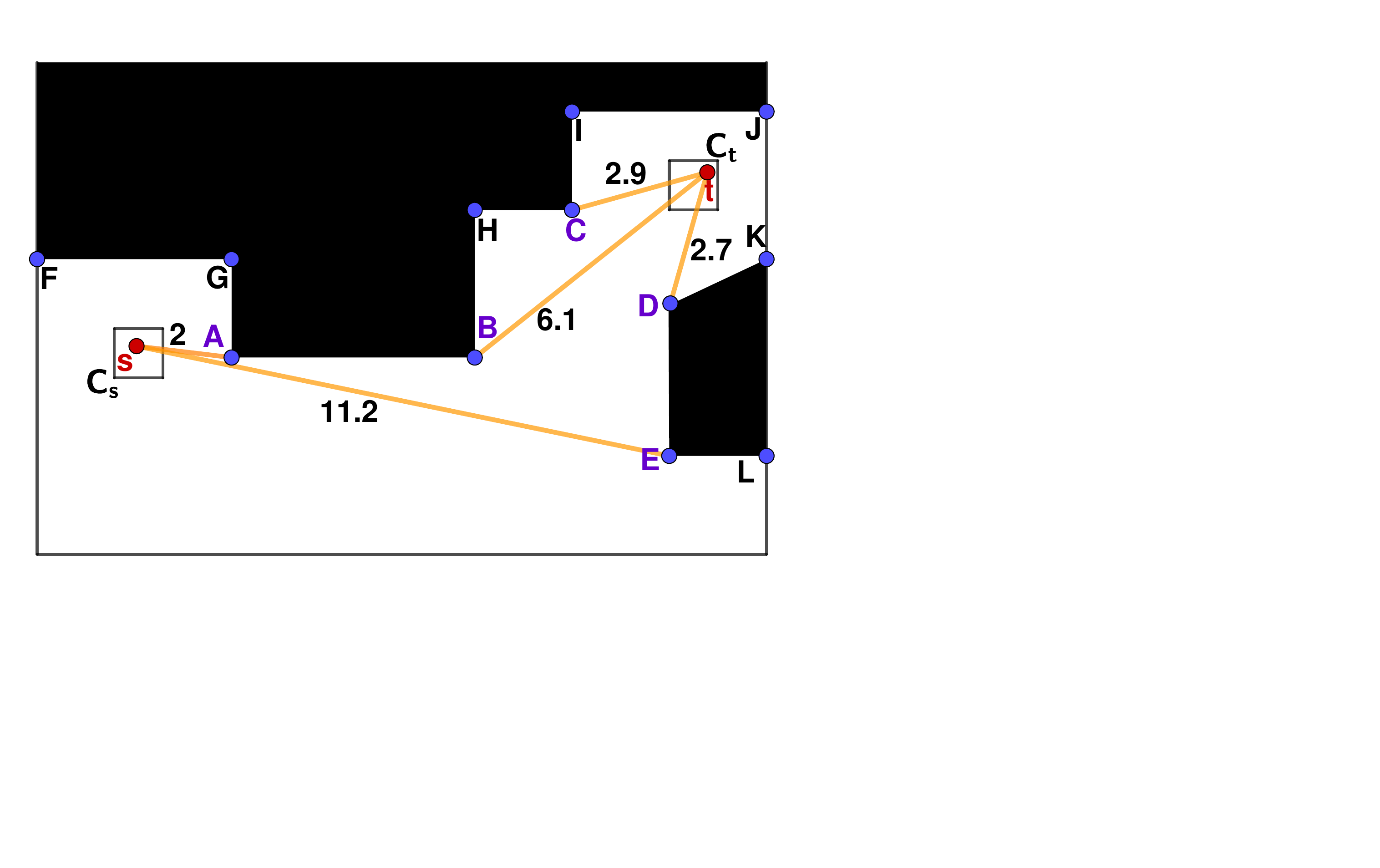}
\caption{Euclidean plane with polygonal obstacles}
\label{fig::running_example}
\end{figure}

\begin{example}
    Table~\ref{tab:vialabels} presents the via-labels for two grid cells, $c_s$ and $c_t$, as illustrated in Figure~\ref{fig::running_example}. To compute these via-labels, the EHL algorithm first identifies the visible convex vertices from each cell, forming visibility lists $L_{c_s} = \{A,B,E\}$ and $L_{c_t} = \{B,C,D\}$. Consider vertex $A$, with its hub label $(B,5.1) \in H(A)$ as shown in Table~\ref{tab:hublabels}. Since $A$ is visible from $c_s$, a via-label $B:(A,5.1)$ is added to indicate that there is a path from $c_s$ to $B$ via $A$, with the shortest distance $d(A,B) = 5.1$. The visibility list $L_{c_s}$ includes $A$, $B$, and $E$, so the hub labels of these vertices are used to insert the via-labels for $c_s$: $A:(A,0)$, $B:(A,5.1)$, $E:(A,10)$, $B:(B,0)$, $B:(E,6.1)$, $D:(E,5.3)$, and $E:(E,0)$. Similarly, the via-labels for $c_t$ are generated.
\end{example}

\subsection{Online Query Processing}
First, we introduce minimal via-distance and how EHL computes it using the stored labels.

\subsubsection{Minimal Via-Distance}
The via-distance $vdist(p,v_j,h_i)$ represents the length of the shortest path between a point $p$ and $h_i$ passing through a convex vertex $v_j$ visible from $p$. Given a via label $h_i\rt(v_j,d_{ij})\in VL_{h_i}(c)$ and a point $p\in c$, if $v_j$ is visible from $p$ then $vdist(p,v_j,h_i)=Edist(p,v_j) + d_{ij}$ where $Edist(p,v_j)$ is the Euclidean distance between $p$ and $v_j$. If $v_j$ is not visible from $p$, we assume $vdist(p,v_j,h_i)=\infty$.
Given a point $p\in c$ and the via labels $VL_{h_i}(c)$, we define the minimum via-distance $vdist_{min}(p,h_i)$ between $p$ and $h_i$ as:


\begin{equation}
vdist_{min}(p,h_i)=\argmin_{h_i\rt(v_j,d_{ij})\in VL_{h_i}(c)} vdist(p,v_j,h_i)
\end{equation}


\begin{table}[t]

    \begin{minipage}[t]{.50\linewidth}
      \centering
      \small
      \setlength{\tabcolsep}{1pt} 
\begin{tabular}{| c | l | }
\hline
 $\bm{H(c_s)}$  & \textbf{Via Labels $\bm{VL_{h_i}(c_s)}$}  \\
 \hline \hline
 A$\rt$ & (A, 0)\\ \hline
 B$\rt$ & (A, 5.1), (B, 0), (E, 6.1) \\ \hline
 D$\rt$ & (E, 5.3) \\ \hline 
 E$\rt$ & (A, 10), (E, 0) \\ \hline
\end{tabular}
    \end{minipage}%
    \begin{minipage}[t]{.50\linewidth}
      \centering    
      \small
      \setlength{\tabcolsep}{1pt} 
      \vspace{-7.25ex} 
\begin{tabular}{| c | l | }
\hline
 $\bm{H(c_t)}$  & \textbf{Via Labels $\bm{VL_{h_i}(c_t)}$}  \\
 \hline \hline
B$\rt$ & (B, 0), (C, 5.1), (D, 5.4)  \\ \hline
C$\rt$ & (C, 0) \\
\hline
D$\rt$ & (D, 0) \\ \hline
\end{tabular}
    \end{minipage} 
    \caption{Via-labels for $c_s$ and $c_t$ shown in Figure~\ref{fig::running_example}}
    \label{tab:vialabels}
\end{table}

\subsubsection{Computing Shortest Distance} 
Let $c_s$ and $c_t$ be the grid cells containing $s$ and $t$, respectively. If $s$ and $t$ are co-visible, then the shortest distance $d(s,t)$ is simply the Euclidean distance between them (i.e., $d(s,t) = Edist(s,t)$). If they are not co-visible, $d(s,t)$ is computed as follows:

\begin{equation}\label{eq:d_s_t}
d(s,t) = \argmin_{h_i\in H(c_s)\cap H(c_t)} vdist_{min}(s,h_i) + vdist_{min}(t,h_i)
\end{equation}

Once $d(s,t)$
is determined, the shortest path $sp(s,t)$ can be retrieved using the successor nodes. Please see~\cite{EHL} for details.

\begin{example}
In Table~\ref{tab:vialabels}, 
$H(c_s)$ and $H(c_t)$ share two common hub nodes, B and D. When B is identified, the $vdist_{min}(s,B)$ is computed using each via-label: $vdist(s,A,B)=EDist(s,A)+d(A,B)=2+5.1=7.1$ and $vdist(s,E,B)=EDist(s,E)+d(E,B)=11.2+6.1=17.3$. The label $B:(B,0)$ is ignored since B is not visible from $s$, resulting in $vdist_{min}(s,B)=vdist(s,A,B)=7.9$. Similarly, $vdist_{min}(t,B)=6.1$ is computed, and the distance is updated to $dist=vdist_{min}(s,B)+vdist_{min}(t,B)=14$. For hub node D, $vdist_{min}(s,D)+vdist_{min}(t,D)=16.5+2.7=19.2$. The algorithm returns the smaller value $d(s,t)=14$.
\end{example}

\section{Euclidean Hub Labeling Star (EHL*)}

In this section, we introduce EHL*, a memory-budgeted version of EHL. EHL* enhances EHL by introducing a compression phase during the offline preprocessing stage. 


\subsection{Offline Preprocessing}

The query performance of EHL depends on the number of labels stored in $c_s$ and $c_t$ -- the fewer, the better. Thus, EHL performs better when the grid cells are small because smaller cells contain fewer labels. However, memory usage increases with smaller cells due to the likelihood of the same labels being stored in multiple cells.  Memory usage in EHL can be significantly reduced without greatly impacting query performance by merging cells, as long as the number of labels per cell does not increase substantially. EHL* aims to achieve this as we detail next.

Recall that each via-label stored in a grid cell $c$ is essentially a hub label copied from a convex vertex $v$, from which at least part of $c$ is visible. As a result, neighboring cells often share many of the same hub labels. By merging adjacent cells that have a high overlap in hub labels, memory usage can be reduced without significantly impacting performance. EHL* leverages this insight by introducing a compression phase that iteratively checks adjacent grid cells and merges those with a high degree of hub label similarity.



\begin{algorithm2e}[t]
\small
\caption{EHL*: Compression Phase}
    \label{algo:ehl}
\setcounter{AlgoLine}{0}
    \SetKwInput{Input}{Input}
    \SetKwInput{Initialization}{Initialization}
    \SetKwInput{Output}{Output}
    \Input{EHL: Euclidean Hub Labeling, $\mathcal{B}$: memory budget.}
    \Output{EHL*: the compressed EHL. M: a mapper that maps each grid cell to the merged region of EHL*.}
    \texttt{initializeScores}(EHL) \label{ehl:assignValue}\;
    Initialize a min heap $H$ with each grid cell of EHL; \label{ehl:minheap} \\
    Initialize a mapper $M$ that maps each grid cell to its corresponding element in min heap $H$; \label{ehl:mapper} \\
    \While {MemoryUsage $> \mathcal{B}$ and $|H|>1$ }
    {\label{ehl:while}
        deheap an element $e$ from $H$ \label{ehl:deheap} \;
        $r \leftarrow$ \texttt{adjacentRegionSelection}($e$,$M$) \; \label{ehl:selectGrid}
            $e \leftarrow$ $e\cup r$; \tcp{merge $r$ into $e$} \label{ehl:merge}
            update $M$ by mapping each cell $c \in r$ to $e$ \; \label{ehl:updateMapper}
            remove $r$ from $H$ and insert $e$ in $H$ \;\label{ehl:update}
    }
    \textbf{return} EHL* and $M$\label{ehl:return}\;
\end{algorithm2e}

Algorithm~\ref{algo:ehl} details the compression phase. From this point on, we use ``region'' to refer to either a single grid cell or the shape resulting from merging two or more cells. The algorithm receives the constructed EHL and a memory budget $\mathcal{B}$ as input. It begins by calling the function \texttt{initializeScores}(EHL) to assign a score to each grid cell in the EHL (line~\ref{ehl:assignValue}). These scores help determine which regions to merge in each iteration. We will explain the score computation in the next section.
Next, the algorithm initializes a min heap $H$ by inserting each grid cell along with its corresponding score (line~\ref{ehl:minheap}) and sets up a mapper $M$ that links each grid cell to its region in $H$ (line~\ref{ehl:mapper}). This mapper allows for efficient tracking of the merged regions associated with each grid cell.

Once initialization is complete, the algorithm begins compressing the EHL through a while loop. In each iteration, the algorithm extracts the top element $e$ from the heap $H$ (line~\ref{ehl:deheap}) and calls the function \texttt{adjacentRegionSelection}$(e, M)$, which examines each adjacent region of $e$ and chooses a region $r$ to merge with $e$ (line~\ref{ehl:selectGrid}). Two regions are considered adjacent if they share a boundary. The mapper $M$ is used to efficiently identify $e$'s adjacent regions. Specifically, a region $e'$ is adjacent to $e$ if a cell $c$ adjacent to $e$ belongs to $e'$ (as determined by the mapper $M$). The criteria for selecting the region $r$ are discussed later. Once the region $r$ is selected for merging with element $e$, they are merged at line~\ref{ehl:merge} as follows .

\begin{enumerate}
    \item The region of element $e$ is expanded to include $r$.
    \item To maintain the correctness of the algorithm, the via-labels of $r$ and $e$ are merged by taking their union. Specifically, for each via-label $h_i\rt(v_j, d_{ij})$ of $r$, we add it to $VL(e)$ if $h_i\rt(v_j, d_{ij}) \notin VL(e)$; otherwise, we ignore it.
    \item The score of $e$ is incremented by the score of $r$, i.e., $s(e) = s(e) + s(r)$, where $s(x)$ denotes score of $x$.
\end{enumerate}

After merging region $r$ into region $e$, the algorithm updates the mapper $M$ by reassigning each grid cell $c \in r$ to the expanded region $e$, ensuring that adjacent regions are correctly tracked in future iterations (line~\ref{ehl:updateMapper}). The region $r$ is then removed from the heap $H$ to prevent redundant merges, and the expanded region $e$ is reinserted into $H$ with its updated score (line~\ref{ehl:update}). This process repeats until the memory usage is less than or equal to the memory budget $\mathcal{B}$ (line~\ref{ehl:while}), at which point EHL* has been successfully constructed within the given budget. The algorithm also halts if all cells have merged into a single region (i.e., the heap contains only one element), which indicates that EHL* cannot be constructed within the memory limit. In our experiments, this situation only arises for certain small maps when the memory budget is set to just $1\%$ of the original EHL memory, which is already quite limited.


The algorithm initially calculates the total memory usage, and throughout the iterations, it tracks the reduction in memory by subtracting the amount saved through label merging. The final output of the algorithm includes the compressed EHL (i.e., EHL*) and the updated mapper $M$, which enables faster query processing as discussed later (line~\ref{ehl:return}).



\subsubsection{Initializing Scores}
The scores assigned to grid cells play a crucial role in the compression phase, as cells with lower scores are more likely to be merged. These scores can be tailored based on the application's specific needs. For example, if certain areas of the map require highly efficient query runtimes, the cells in those areas can be given higher scores to reduce the likelihood of being merged, thereby retaining fewer labels and improving performance. In the absence of such requirements, EHL* assigns the same score to all cells, i.e., $s(c)=1$ for each cell. This simple approach ensures that the merged regions have approximately equal sizes and similar number of hub labels, leading to relatively consistent query performance across different map areas.

\subsubsection{Adjacent Region Selection} 

To reduce memory usage without significantly affecting query runtime performance, the key is to keep the number of labels per region as small as possible. This can be accomplished by merging regions with substantial overlap in their hub labels. EHL* does this by examining all adjacent regions $R$ of $e$ and selecting the region $r \in R$ whose hub labels exhibit the highest Jaccard similarity with those of $e$ (ties are broken arbitrarily).
\begin{equation}\label{eq:sim}
r = \arg\max_{r' \in R} \left( \frac{|H(r') \cap H(e)|}{|H(r') \cup H(e)|} \right)
\end{equation}

\begin{figure}[t]
\centering
\includegraphics[width=0.8\columnwidth]{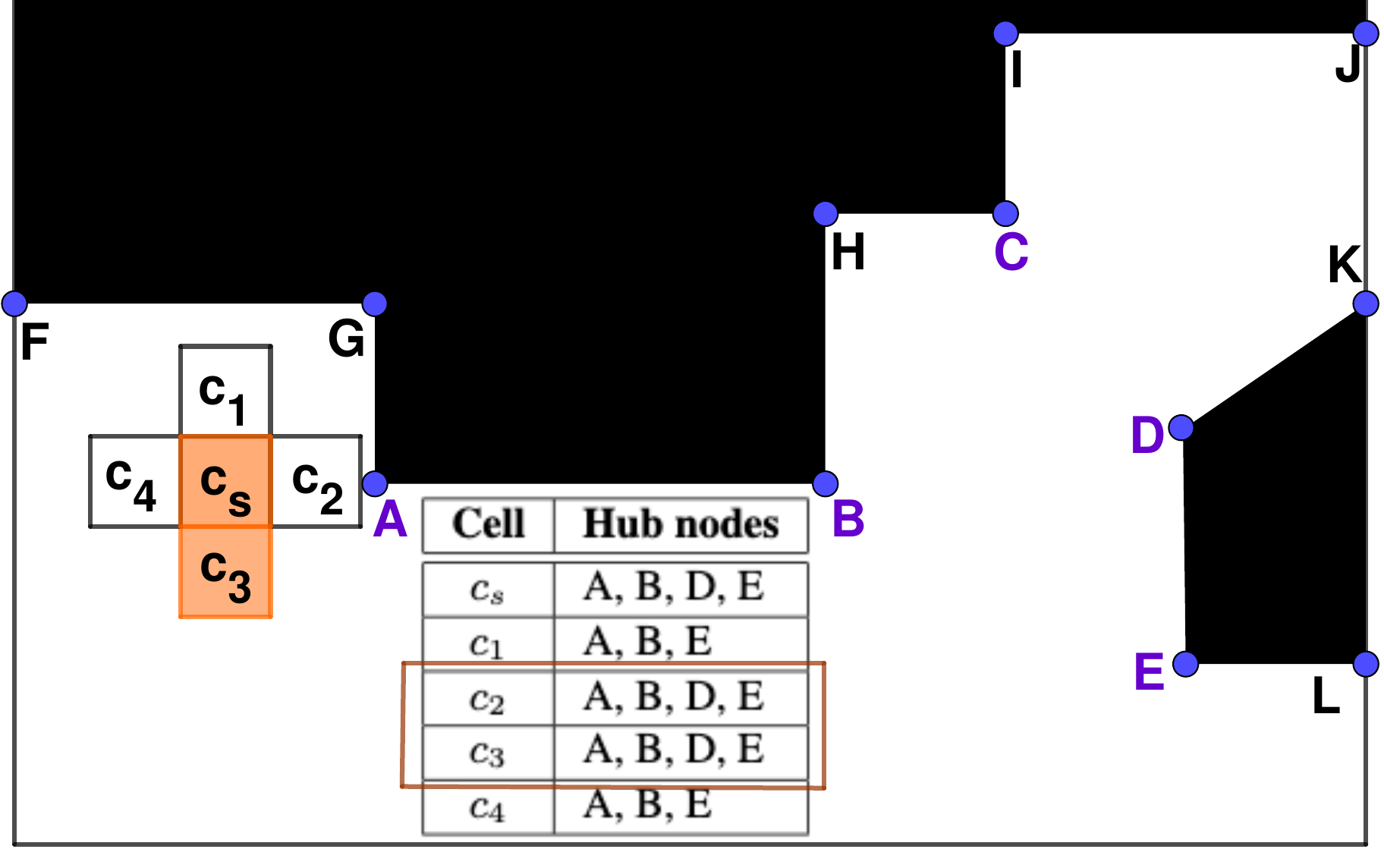}
\caption{Grid cell $c_s$ with its adjacent neighbors}
\label{fig::merge}
\end{figure}

\begin{example}
    Figure~\ref{fig::merge} shows a grid cell $c_s$ that needs to be merged with one of its four neighboring cells: $c_1$ to $c_4$. The table in the figure lists the hub nodes for each of these cells. Cells $c_1$ and $c_4$ have a Jaccard similarity of 0.75 with $c_s$, while both $c_2$ and $c_3$ have a similarity of 1. $c_2$ and $c_3$ have the highest similarity and we arbitrarily select $c_3$ for merging with $c_s$. Table~\ref{tab:mergevia} shows the via labels for $c_s$ after the merge with $c_3$. Two new via labels (highlighted in red) from $c_3$ are added to the label set of $c_s$ as a result of the merge.
\end{example}

\renewcommand{\arraystretch}{1.1}
\begin{table}[t]
    \centering \small
\begin{tabular}{| c | l | }
\hline
 $\bm{H(c_s)}$  & \textbf{Via Labels $\bm{VL_{h_i}(c_s)}$}  \\
 \hline \hline
 A$\rt$ & (A, 0)\\ \hline
 B$\rt$ & (A, 5.1), (B, 0), (E, 6.1), \textcolor{red}{(D, 5.4)} \\ \hline
 D$\rt$ & (E, 5.3), \textcolor{red}{(D, 0)} \\ \hline 
 E$\rt$ & (A, 10), (E, 0)
 \\ \hline
\end{tabular}
    \caption{Via-labels for $c_s$ after merging $c_3$ into it. Newly inserted labels are shown in red.}
    \label{tab:mergevia}
\end{table}
\renewcommand{\arraystretch}{1}

\subsection{Online Query Processing}
EHL stores labels in uniform grid cells whereas EHL* differs by merging these cells into arbitrary-shaped regions. Despite this, both EHL and EHL* maintain the via-label list $VL(e)$ within each region $e$, so the query processing algorithm remains the same. 
Note that EHL* needs to first locate the regions $e_s$ and $e_t$ that contain $s$ and $t$, respectively. This can be easily done in $O(1)$ by identifying the grid cells that contain $s$ and $t$ and then using the mapper $M$ to find the corresponding regions $e_s$ and $e_t$.

\subsection{Query Workload-Aware EHL*}

In many real-world scenarios, such as computer games, query workload can often be predicted because certain regions of a map are visited more frequently due to gameplay mechanics, player behavior patterns, or specific objectives. \emph{Workload-Aware algorithms}~\cite{sheng2023wisk} can leverage such knowledge if, for example, expected query distribution is known or can be predicted. 
We adapt EHL* for workload-aware scenarios by modifying its scoring computation and adjacent region selection.

\subsubsection{Initializing Scores}
If the query distribution is known or can be predicted, EHL* can improve performance by keeping regions with higher expected workloads smaller, which in turn reduces the number of labels and enhances performance. To achieve this, EHL* prioritizes merging regions with lower expected query activity. Let $w_c$ represent the expected workload of a cell $c$, defined as the anticipated number of queries where $s$ or $t$ falls within $c$. EHL* assigns an initial score to each cell as $s(c) = 1 + w_c$, ensuring that no cell has a zero score, and that cells with higher workloads have higher scores, making them less likely to be merged.

\subsubsection{Adjacent Region Selection} 
When merging regions, the primary goal remains minimizing the number of labels. Thus, EHL* still prioritizes similarity when selecting adjacent regions for merging. However, instead of solely considering similarity, we also factor in the expected workload, represented by the scores $s(r)$ for each region $r$. The selection of an adjacent region $r \in R$ is now determined by combining these two criteria:


\begin{equation}\label{eq:sim2}
r = \arg\max_{r' \in R} \left( (1 - \alpha) \left( \frac{|H(r') \cap H(e)|}{|H(r') \cup H(e)|} \right) + \alpha \left( \frac{1}{s(r')} \right) \right)
\end{equation}

Note that regions with higher workloads $s(r')$ are less likely to be selected for merging. The weighting factor $\alpha$, set to 0.2 based on experimental results, balances the influence of similarity and workload.

\section{Experiments}

 \renewcommand{\arraystretch}{1}
\begin{table}[t]
    \centering \small
\begin{tabular}{| l | c | c | c | c | }
\hline
\textbf{ }  & \textbf{\#Maps} & \textbf{\# Queries} & \textbf{\#Vertices} & \textbf{\#CV} \\ \hline 
DAO & 156 & 159,464 & 1727.6 & 926.5 \\ \hline
DA & 67 & 68,150 & 1182.9 & 610.8 \\ \hline
BG & 75 & 93,160 & 1294.4 & 667.7 \\ \hline
SC & 75 & 198,224 & 11487.5 & 5792.7 \\ \hline
\end{tabular}
    \caption{
    Total \# of maps and queries, and average \# of vertices and convex vertices (\#CV) in each benchmark.
    }
    \label{tab:benchmark}
\end{table}









\begin{table*}[t!]
\setlength{\tabcolsep}{5pt} 
    \centering \small
\begin{tabular}{ c| c| c | r r r r r r |  r r r r r  }
\toprule
 \multirow{2}{*}{Map} & \multicolumn{2}{c|}{\multirow{2}{*}{Query Set}}  & \multicolumn{6}{c|}{EHL*}& \multicolumn{5}{c}{Competitors}
  \\\cline{4-14}
    & \multicolumn{2}{c|}{} & 80\% & 60\% & 40\% & 20\% & 10\%& 5\% 
    &  EHL-1 & EHL-2 & EHL-4 & EPS & Polyanya

    \\
\midrule
\midrule

 \multirow{6}{*}{DAO}&
\multicolumn{2}{c|}{Memory(MB)}
    & 105.5 & 78.8 & 52.4 & 26.3 & 13.6 & 7.5 & 134.8 & 57.8 & 27.6 &  0.06 & -  
    \\\cline{2-14} & 
\multicolumn{2}{c|}{Build Time(Secs)}
    & 6.63 & 7.21 & 7.98 & 8.99 & 9.43 & 9.68 & 4.57 & 1.24 & 0.36 & 0.02 & -  
    \\\cline{2-14}
& \multirow{4}{*}{\rotatebox{90}{Query($\mu$s)}} &Unknown
  &2.02 & 2.08 & 2.20 & 2.49 & 3.12 & 4.40 & 1.84 & 2.28 & 3.19 & 25.94 & 176.50
   \\\cline{3-14 }
&&Cluster-2
 &1.00	&0.99 &0.98 &0.98  & 1.03 &1.15	&0.98 &1.27 &1.88 & 7.22 & 17.51
 \\\cline{3-14 }
 
&&Cluster-4
 & 1.22 & 1.21 & 1.20 & 1.23 & 1.33 & 1.59& 1.21 & 1.55 & 2.23 & 10.01 & 32.66
 \\\cline{3-14 }
 
&&Cluster-8
 & 1.35 & 1.34 & 1.35 & 1.43 & 1.58 & 2.00 & 1.33 & 1.69 & 2.43 & 11.69 & 39.60
  \\\hline
   
 \multirow{5}{*}{DA}&
\multicolumn{2}{c|}{Memory(MB)}
& 36.0 & 27.0 & 18.2 & 9.4 & 5.1& 3.0  & 47.2 & 19.6 & 9.0 & 0.21 & - 
    \\\cline{2-14}
&\multicolumn{2}{c|}{Build Time(Secs)}
    & 2.76 & 2.88 &3.05 & 3.26 & 3.37 & 3.43 & 1.67 & 0.51 & 0.18 & 0.11 & -  
    \\\cline{2-14}
& \multirow{4}{*}{\rotatebox{90}{Query($\mu$s)}} &Unknown
 &1.64 & 1.68 & 1.75 & 1.93 & 2.25 & 2.86 & 1.54 & 1.85 & 2.48 & 12.17 & 37.85
  \\\cline{3-14}
&&Cluster-2
 & 0.89 & 0.88 & 0.90 & 0.91 & 0.93 & 1.00 & 0.88 & 1.09 & 1.47 & 5.29 & 9.45 
 \\\cline{3-14}
 
&&Cluster-4
 &1.10 & 1.09 & 1.12 & 1.14 & 1.20 & 1.38 & 1.09 & 1.33 & 1.82 & 7.13 & 14.10
 \\\cline{3-14 }
 
&&Cluster-8
 &1.24 & 1.24 & 1.28 & 1.33 & 1.46 & 1.79 & 1.23 & 1.52 & 2.07 & 7.89 & 14.37
  \\\hline
   
 \multirow{5}{*}{BG}&
\multicolumn{2}{c|}{Memory(MB)}
& 150.8 & 112.8 & 75.0 & 37.6 & 19.3 & 10.3 & 193.0 & 71.2 & 31.8 & 0.12 & - 
    \\\cline{2-14}
&\multicolumn{2}{c|}{Build Time(Secs)}
    & 10.13 & 10.92 &11.85 & 13.04 & 13.72 & 14.00 & 6.91 & 1.80 & 0.51 & 0.04 & -  
    \\\cline{2-14}
& \multirow{4}{*}{\rotatebox{90}{Query($\mu$s)}}&Unknown
 & 1.34 & 1.37 & 1.43 & 1.55 & 1.79 & 2.25 & 1.27& 1.48 & 1.93 & 9.77 & 13.25
  \\\cline{3-14}
&&Cluster-2
 &0.75 & 0.74 & 0.78 & 0.80 & 0.79 & 0.82 & 0.73 & 0.85 & 1.11 & 4.32 & 6.84
 \\\cline{3-14}
 
&&Cluster-4
 &0.96 & 0.97 & 1.03 & 1.02 & 1.04 & 1.11 & 0.95 & 1.10 & 1.42 & 6.32 & 10.12
 \\\cline{3-14 }
 
&&Cluster-8
 &1.09 & 1.11 & 1.17 & 1.19 & 1.23 & 1.37 & 1.08 & 1.23 & 1.59 & 7.29 & 11.66
  \\\hline
   
 \multirow{5}{*}{SC}&
\multicolumn{2}{c|}{Memory(MB)}
 &2060.7 & 1538.5 & 1017.3 & 503.9 & 253.6 & 130.9 & 2586.9 & 1096.4 & 522.5 & 2.3 & -
 \\\cline{2-14}
&\multicolumn{2}{c|}{Build Time(Secs)}
    & 178.96 & 189.44 &207.40 & 226.60 & 241.39 & 243.18 & 118.06 & 30.62 & 8.01 & 3.49 & -  
    \\\cline{2-14}
& \multirow{4}{*}{\rotatebox{90}{Query($\mu$s)}}&Unknown
 & 3.43 & 3.50 & 3.96 & 4.29 & 5.05 & 6.47 & 3.34 & 4.13 & 5.40 &68.20 &  261.16
  \\\cline{3-14}
&&Cluster-2
 &1.83 & 1.93 & 1.79 & 1.94 & 1.81 & 1.89 & 1.92 & 2.22 & 2.93 & 30.56 & 83.23
 \\\cline{3-14 }
 
&&Cluster-4
 &2.48 & 2.43 & 2.49 & 2.41 & 2.43 & 2.63 & 2.57 & 2.96 & 3.89 & 43.74 & 104.74
 \\\cline{3-14 }
 
&&Cluster-8
 &2.95 & 2.97 & 3.02& 3.19 & 3.06 & 3.47 & 3.08 & 3.59 & 4.66 & 51.94 & 120.16
  \\\hline

\bottomrule

\end{tabular}
    \caption{Memory, build time and runtime comparison between EHL* and the competitors for different scenarios.}
    \label{tab:runtime}
\end{table*}

\subsection{Settings}

We run our experiments on a 3.2 GHz Intel Core i7 machine with 32GB of RAM. The algorithms are all implemented in C++ and compiled with -O3 flag. Following existing studies, we conduct experiments on the widely used game map benchmarks~\cite{gridBenchmarks}, which consist of four games: Dragon Age II (DA), Dragon Age: Origins (DAO), Baldur's Gate II (BG), and StarCraft (SC). In total, we have 373 game maps from the four game benchmarks, each of which is represented
as a grid map. 
Table~\ref{tab:benchmark} provides details of the benchmarks.

\noindent\textbf{Queries.}
To represent scenarios where the query distribution is unknown, denoted as \textbf{Unknown}, we use the original dataset provided for each game map~\cite{gridBenchmarks}.   To simulate known query distributions, we generate synthetic query sets labeled \textbf{Cluster-x}, where $x$ indicates the number of clustered regions (2, 4, or 8) within a map. These clusters are represented by rectangles, each generated by selecting a random central point within the traversable area of the map and ensuring the rectangle does not extend beyond map boundaries. The rectangles, with side lengths set to 10\% of the width and height of the map, are positioned to ensure at least one other rectangle is reachable, preventing isolation. We generate source $s$ and target $t$ pairs as randomly selected locations within these rectangles ensuring that the path between $s$ and $t$ exists for each generated query.
Using this approach, we generate (i) 500,000 $s$ and $t$ pairs as historical queries on the map, which EHL* uses to calculate the workload for each cell and build the index, and (ii) 2,000 $s$ and $t$ pairs to evaluate runtime performance.
To evaluate runtime, each query is run five times, and the average is reported.


\noindent\textbf{Algorithms Evaluated.} 
We compare our approach with the state-of-the-art algorithm\footnote{https://github.com/goldi1027/EHL}, EHL-x~\cite{EHL}, where 
x denotes the cell size in EHL, varying x from 1 to 4 (similar to the original work). Additionally, we compare our method with Polyanya\footnote{https://bitbucket.org/mlcui1/polyanya}~\cite{polyanya}, an optimal online pathfinding algorithm that runs on a navigation mesh, and EPS\footnote{https://github.com/bshen95/End-Point-Search}~\cite{EPS}, an optimal offline pathfinding algorithm that utilizes a Compressed Path Database 
 (CPD)~\cite{fast_cpd}. 
We evaluate EHL* under different memory budgets, denoted as EHL*-x, where x represents the percentage of memory used compared to EHL-1, e.g., the budget for EHL*-60 is 60\% of the memory used by EHL-1. For reproducibility, implementation of EHL* is available  online\footnote{https://github.com/goldi1027/EHL*}.

\begin{figure*}[t]
\scriptsize
\begin{tabular}{@{~}llll@{~}}
\begin{minipage}{.24\linewidth}
\centering
\includegraphics[scale=0.367]{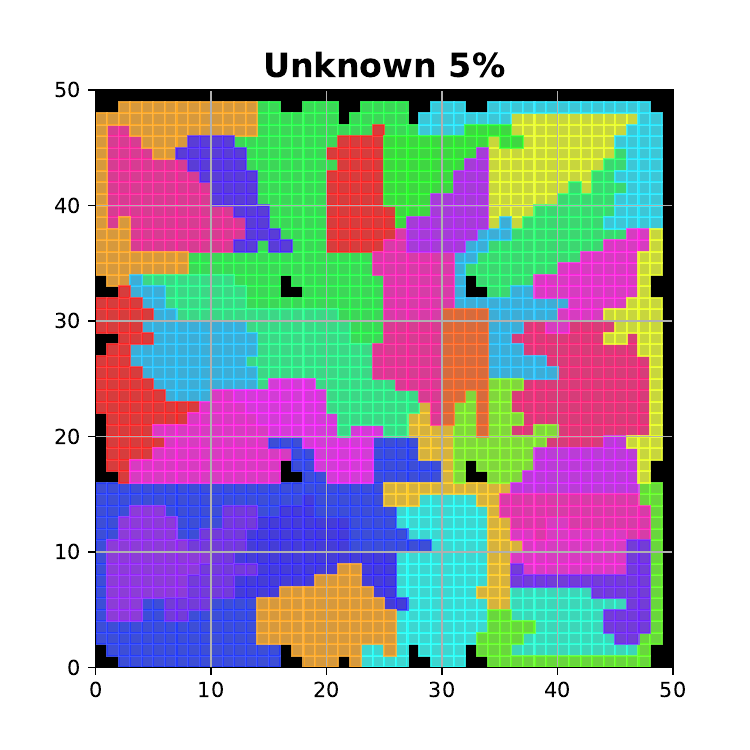}
\end{minipage}

\begin{minipage}{.24\linewidth}
  \centering
\includegraphics[scale=0.367]{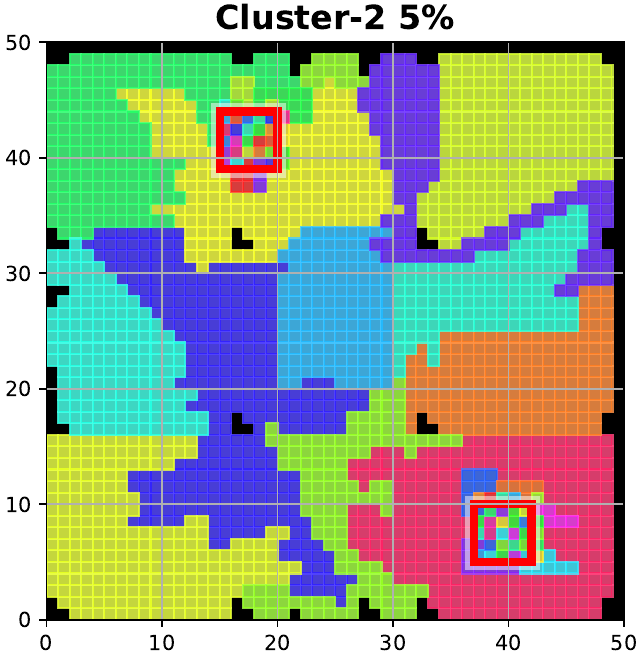}
\end{minipage}&

\begin{minipage}{.24\linewidth}
\centering
\includegraphics[scale=0.367]{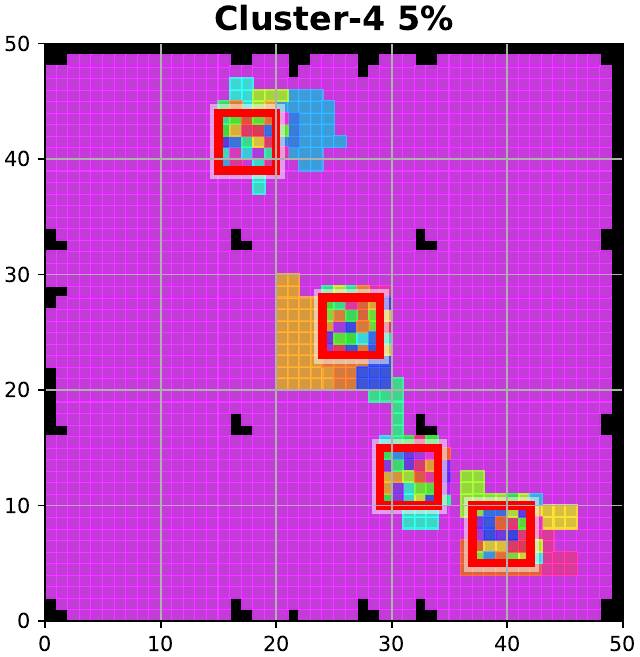}

\end{minipage}

\begin{minipage}{.24\linewidth}
\centering
\includegraphics[scale=0.367]{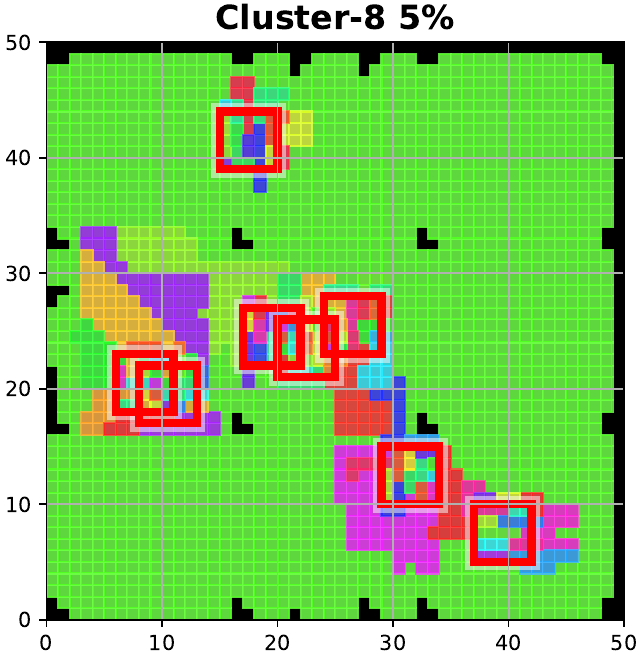}
\end{minipage}

\end{tabular}
\caption{Distribution of merged grids for arena map for different cluster scenarios at 5\% memory. The red rectangles represents the cluster regions generated in the arena map. Black polygons are the obstacles.}
\label{fig::grid_distribution}
\end{figure*}

\subsection{Experiment 1: Preprocessing Time and Space} 
Table~\ref{tab:runtime} compares the average memory usage (in MB) and build time (in secs) of EHL* for different memory budgets with EHL, EPS and Polyanya. 
Compared to EHL, EHL* offers precise memory usage control, successfully reducing memory down to 5\%, making it adaptable to different devices based on application needs. While EHL also offers a memory-runtime trade-off by increasing cell sizes, its memory usage cannot be predicted before index construction and lacks fine-tuned control.
EPS and Polyanya, however, require significantly less memory but at the expense of a much higher query runtime as shown shortly. Polyanya's memory usage is not shown, as it is an online algorithm that requires no index, aside from a navigation mesh with a negligible memory footprint. 
Although EHL* requires nearly twice the preprocessing time compared to EHL-1 and significantly more preprocessing time  than the other competitors, the overall build time remains reasonable and practical for most application needs.

\subsection{Experiment 2: Memory-Runtime Tradeoff} 
Next, we discuss the memory-runtime tradeoff for each algorithm. Table~\ref{tab:runtime} also shows 
the query runtime (in $\mu$s) for both the unknown and clustered queries.
EPS and Polyanya offer lower memory usage but are 1-2 orders of magnitude slower than EHL and EHL*. EHL* reduces memory usage of EHL-1 from 80\% to 20\% with minimal impact on its runtime, but further reductions to 10\% and 5\% lead to noticeable increases in query time due to significantly larger merged regions  leading to slower runtimes. 
Comparing with EHL on unknown query distributions, EHL*-80 and EHL*-60 use less memory than EHL-1 while keeping runtimes competitive, and EHL*-40 and EHL*-20, while similar in memory usage to EHL-2 and EHL-4, deliver better runtimes, thus outperforming EHL.

When the query distribution is known in advance, EHL* leverages this information to significantly improve the performance. 
With a small number of cluster regions (e.g., Cluster-2 and Cluster-4), EHL* can reduce the memory budget from 80\% to 5\% while maintaining similar query runtimes, making the memory reduction nearly cost-free. Compared to EHL, EHL* remains competitive with EHL-1 across all memory budgets and consistently outperforms EHL-2 and EHL-4, demonstrating a superior memory-runtime tradeoff.
As the number of cluster regions increases (e.g., Cluster-8), memory reduction becomes more challenging, leading to a more noticeable increase in query runtime. Nevertheless, the performance remains better than EHL for Cluster-8, e.g., when the memory budget is reduced to 5\%, EHL*-5 still outperforms EHL-4 which consumes 3-5 times more memory.

Figure~\ref{fig::grid_distribution} illustrates how EHL* merges regions when the memory budget is set to 5\%. The merged regions are displayed in different colors for various query sets on the arena map from the DA benchmark. When the query distribution is unknown, EHL* combines grid cells into regions of approximately equal size to efficiently handle queries that may arise at any random position.
However, when the query distribution is known, EHL* prioritizes merging cells outside the cluster regions, leaving more granular grid cells within clusters to efficiently answer queries. This behaviour becomes clearer as the number of cluster regions increases. When the number of clusters is large (e.g., Cluster-8), merging cells outside the clusters may not satisfy the memory budget, forcing some merging within the clusters which leads to slower runtimes for the queries within the clusters.

\begin{table}[t!]
\setlength{\tabcolsep}{0.8pt} 
\renewcommand{\arraystretch}{1.12}
    \centering \small
\begin{tabular}{ c| c | r r r | r r r |  r r r  }
\toprule
 \multirow{2}{*}{Dist} & \multirow{2}{*}{D} & \multicolumn{3}{c|}{EHL* (Known)}& \multicolumn{3}{c|}{EHL* (Unknown)} & \multicolumn{3}{c}{Competitors}
  \\\cline{3-11}
    & &  80\% & 40\% & 20\% & 80\% & 40\% & 20\% 
    &  EHL-1 & EHL-2 & EHL-4

    \\
\midrule
\midrule

 \multirow{3}{*}{100\%}
&C-2
 & 1.30	& 1.33 & 1.34 & 1.45 & 1.56  & 1.79 & 1.30 & 1.69 & 2.35
 \\\cline{3-11 }
 
&C-4
 & 1.44 &  1.48 & 1.50 & 1.58 & 1.76 & 1.87 & 1.45 & 1.79 & 2.50
 \\\cline{3-11 }
 
&C-8
 & 1.53 & 1.48 & 1.64 & 1.74 & 1.82 & 2.04  &1.51  & 1.92 &  2.64
    \\\hline
   
 \multirow{3}{*}{80\%}
&C-2
 & 1.48  & 1.54 & 1.64 & 1.54 & 1.65 & 1.75 & 1.38  & 1.75 & 2.41
 \\\cline{3-11 }
 
&C-4
 & 1.50 & 1.55 & 1.86 & 1.71 & 1.82 & 1.89 & 1.47 & 1.83 & 2.53
 \\\cline{3-11 }
 
&C-8
 & 1.63 & 1.73 & 2.03 & 1.75 & 1.87 & 1.97 & 1.54 & 1.91 & 2.63
  \\\hline
   
 \multirow{3}{*}{50\%}
&C-2
 & 1.59 & 1.67 & 1.82 & 1.66 & 1.76 & 1.84 & 1.47 & 1.82 & 2.48
 \\\cline{3-11 }
 
&C-4
 & 1.68 & 1.80 & 2.34 & 1.70 & 1.81 & 1.95 & 1.52 & 1.87 & 2.53
 \\\cline{3-11 }
 
&C-8
 & 1.70 & 1.87 & 2.43 & 1.72 & 1.87 & 2.01 & 1.55 & 1.92 & 2.61
  \\\hline
   
 \multirow{3}{*}{20\%}
&C-2
 & 1.75 & 1.85 & 2.20 & 1.87 & 1.95 & 1.98 & 1.53 & 1.89 &  2.57
 \\\cline{3-11 }
 
&C-4
 & 1.75 & 1.88 & 2.56 & 1.72 & 1.86 & 2.04 & 1.56  & 1.91 & 2.58
 \\\cline{3-11 }
 
&C-8
 & 1.70 & 1.96 & 3.09 & 1.79 & 1.94 & 2.08 & 1.57 & 1.94 & 2.62

  \\\hline

\bottomrule

\end{tabular}
    \caption{Running time comparison between EHL* and competitors for mixed distribution scenarios for DA benchmark.}
    \label{tab:runtime_distribution}
\end{table}
\subsection{Experiment 3: Query Distribution Deviations}
In this section, we analyze the performance of EHL* for queries that diverge from the predicted distribution. We construct EHL* based on the assumed Cluster-x distribution, but only $y$\% of the queries adhere to this distribution, while the remaining $(100 - y)$\% are generated randomly. We vary $y$ from 100 to 20. Table~\ref{tab:runtime} shows the average runtime (in $\mu$s) for the three cluster query sets in the DA benchmark. We compare our approaches, EHL* (known), which tries to exploit query distribution, and EHL* (unknown), which has no information on query distribution, with EHL-1, EHL-2, and EHL-4.
As expected, when a larger number of queries deviate from the predicted distribution (i.e., smaller $y$\%), the performance of EHL* (known) drops significantly, especially when the budget is 20\% of EHL-1 (although still competitive compared to EHL-4). This occurs because EHL* (known) tends to merge grid cells outside the cluster region when memory budget is constrained. As a result, when queries deviate from the cluster regions, larger regions with more via-labels are required to answer the queries, leading to slower query runtimes. In contrast, EHL* (unknown) and EHL remain stable as the number of deviated queries increases since they do not depend on pre-existing query distributions to build index. When most test queries follow the predicted distributions (i.e., 100\% and 80\%), EHL* (known) outperforms EHL* (unknown) and EHL. However, when more queries deviate from the prediction (i.e., 50\% and 20\%), EHL* (known) performs worse than both approaches.

\section{Conclusion}


We propose EHL*, an improved version of EHL that offers the flexibility to build the index within a specified memory budget while optimizing query runtime. 
Experiments show that EHL* reduces memory usage while maintaining competitive query runtimes. Additionally, when query distributions are known in advance, EHL* leverages this information to build an index that outperforms EHL in both memory efficiency and query runtime. Future work includes exploring exploring ML-based regions for further improvements.

\newpage

\bibliography{aaai25}

\end{document}